# Structural Dissolution: How Artificial Intelligence Dismantles Coordination Architecture and Reconfigures the Political Economy of Production

**Author:** Chao Li, Chunyi Zhao **Affiliation:** Founder, AI Edtech Governance Trust; Independent Researcher in AI Governance
**Date:** April 2026 **Status:** Working Paper — Draft for Comment

## Abstract

This paper introduces the *Structural Dissolution Framework* to explain how artificial intelligence systematically dismantles the coordination architecture of traditional industries. We argue that AI systems — by absorbing human multimodal interfaces encompassing language, vision, and behavioral data — dissolve the boundaries that have historically separated firms, markets, experts, and consumers into distinct coordinating agents. This dissolution is not a quantitative efficiency improvement but a qualitative structural transformation producing four fundamental shifts: (1) the erasure of firm, industry, and coordination boundaries; (2) a transition in value creation from physical resources and human collaboration to continuous token flows generated by data refinement loops; (3) the emergence of a new positional logic in which control over domain-specific data refinement infrastructure constitutes control over the new core of production relations; and (4) the emergence of regional data sovereignty entities as the replacement organizational unit that fills the structural space vacated by dissolved firms and markets. The mechanism through which dissolution operates — termed *Interface Internalization* — describes how inter-agent coordination processes are absorbed into intra-system computation. We challenge the Coasian premise that transaction cost minimization determines organizational boundaries, arguing instead that AI renders this boundary economically obsolete. The firm persists as a legal and physical entity but ceases to function as a coordination mechanism, becoming instead a data node within regionally administered AI infrastructure. Using resource-dependent regional economies as an illustrative theoretical case, we demonstrate how this framework explains both the creative dimension of AI adoption — converting seasonal industries into perpetual economic infrastructure — and its destructive dimension — the systematic replacement of intermediate coordination roles and traditional employment structures.

## 1. Introduction

Since Coase's foundational inquiry into why firms exist (Coase, 1937), the dominant framework in organizational economics has held that the boundary between firm and market is determined by the relative costs of coordination. When internal organization is cheaper than market contracting, economic activity is internalized within the firm; when market transactions are cheaper, activity is externalized. Williamson (1975, 1985) formalized this insight into transaction cost economics, which has shaped decades of thinking about industrial organization, corporate strategy, and economic policy.

The emergence of large-scale artificial intelligence systems requires us not to refine this framework but to question its foundational premise. The Coasian question — where should coordination take place, inside firms or across markets — assumes that coordination between distinct agents is a permanent feature of organized economic activity. We challenge this assumption. When AI systems absorb the interfaces through which human coordination operates, they do not merely shift coordination from one institutional locus to another. They dissolve the structural boundaries that made the question meaningful in the first place.

We call this process *Structural Dissolution*: the systematic dismantling of the coordination architecture that has historically organized production across multiple distinct agents — consumers, firms, experts, and governments — into clearly bounded roles and relationships. Structural Dissolution does not produce a new, more efficient version of the old structure. It produces a qualitatively different organizational form in which the old structure's components persist as physical and legal artifacts while losing their defining economic functions.

This paper makes four contributions to the literature.

First, we introduce the *Structural Dissolution Framework* as an overarching theoretical lens for analyzing AI's impact on economic organization. Unlike frameworks that treat AI as a productivity-enhancing technology, the Structural Dissolution Framework treats AI as an organizational force that transforms the architecture of production itself.

Second, we develop *Interface Internalization* as the core mechanism through which Structural Dissolution operates. AI systems capable of processing linguistic, visual, and behavioral data simultaneously absorb the multimodal interfaces through which economic agents have historically coordinated, converting inter-agent coordination into intra-system computation.

Third, we identify four structural transformations produced by this dissolution process and demonstrate their logical interdependence, showing that together they constitute a coherent new economic paradigm rather than a collection of isolated disruptions.

Fourth, we introduce the concept of *regional data sovereignty entities* as the organizational unit that emerges to fill the structural space vacated by dissolved firms and markets, and develop the policy implications of this emergence for regional economic governance.

The remainder of the paper is structured as follows. Section 2 reviews the relevant literature. Section 3 presents the Structural Dissolution Framework and its four transformations. Section 4 develops the Interface Internalization mechanism. Section 5 presents data refinement loops and token fiscal flows as the new value creation and capture mechanism. Section 6 introduces regional data sovereignty and presents an illustrative theoretical case. Section 7 discusses policy implications. Section 8 concludes.

---

## 2. Literature Review

### 2.1 Transaction Cost Theory and the Boundaries of the Firm

Coase (1937) established that firms exist because markets involve costs — costs of discovering prices, negotiating contracts, and enforcing agreements. Williamson (1975, 1985) extended this into a comprehensive framework, identifying asset specificity, uncertainty, and frequency as the key determinants of organizational boundaries.

This tradition shares a foundational assumption we identify as the *coordination permanence assumption*: that coordination between distinct economic agents is a necessary and irreducible feature of organized production. The only question is institutional: which arrangement — firm or market — performs coordination at lower cost. We argue that AI renders this assumption empirically contingent rather than logically necessary. When coordination can be performed by a computational system rather than between agents, the Coasian choice set collapses.

## 2.2 Data as a Factor of Production

Jones and Tonetti (2020) provide the most rigorous treatment of data as a non-rival production factor in mainstream economics. Their framework establishes that because data can be used simultaneously by multiple parties, the socially optimal policy is typically broad data sharing. This has driven a growing literature on data markets, data intermediaries, and the economics of privacy.

The Structural Dissolution Framework departs from this literature in a critical respect. The Jones-Tonetti model treats data as a static input whose value derives from its use at a given point in time. We argue that the distinctive feature of AI-generated data assets is their dynamic and recursive character: data produced in the operation of an AI system feeds back into the system's training, improving performance and enabling richer future data capture. This recursive property — which we formalize as the data refinement loop — generates compounding returns structurally distinct from the diminishing returns of conventional factor inputs.

## 2.3 Platform Economics and Multi-Sided Markets

Rochet and Tirole (2003) established the theoretical foundations for platforms as multi-sided markets that create value by facilitating interactions between distinct user groups. Platform economics has been applied extensively to digital marketplaces and social networks.

AI systems operating through Interface Internalization differ from platforms in a structurally significant way. Platforms mediate between agents who remain distinct; they reduce coordination friction without eliminating coordination as an inter-agent activity. Interface Internalization goes further: it absorbs the agents' coordination functions into the system, such that agents continue to exist but no longer perform the economic function that justified their coordination. This distinction has significant implications for competition policy and industrial organization theory.

## 2.4 Regional Development and the Political Economy of AI

Acemoglu and Restrepo (2019) document AI's labor displacement effects, while Brynjolfsson and McAfee (2014) examine its implications for productivity and inequality. This literature has primarily asked how regions adapt to AI-driven disruption. Our framework poses a complementary question: how can regions *capture* the organizational value that Structural Dissolution creates, by positioning themselves as administrators of the new data sovereignty infrastructure that replaces dissolved coordination structures.

---

# 3. The Structural Dissolution Framework

## 3.1 Core Argument

The Structural Dissolution Framework holds that AI systems, by absorbing human multimodal interfaces, perform a systematic dissolution of the boundaries that structure economic production. These boundaries — between firms and markets, between industries, between the roles of consumer, producer, expert, and intermediary — are not natural features of economic reality but products of the transaction costs and coordination requirements of human-mediated production. When AI absorbs the interfaces through which coordination operates, these boundaries lose their economic rationale and dissolve.

This dissolution is not destruction. The physical assets, legal entities, and human participants that previously operated within these boundaries continue to exist. What dissolves is their structural function: the role they played in coordinating production across boundaries. What remains are components of a new configuration — one organized not around boundary-maintaining coordination but around boundary-transcending computation.

## 3.2 Four Fundamental Transformations

Structural Dissolution produces four interdependent transformations.

**Transformation 1: Erasure of Structural Boundaries**

The boundaries that define conventional economic organization — the boundary of the firm, the boundary of the industry, the boundary between coordinating roles — are products of coordination requirements. Firms exist because bounded internal coordination is cheaper than unbounded market contracting. Industry boundaries exist because expertise and capital are specialized around particular coordination architectures. Role boundaries between consumer, producer, and expert exist because coordination requires agents to occupy defined positions in a transaction.

When AI absorbs coordination functions, each of these boundaries loses its economic foundation. The firm boundary dissolves because there is no longer a coordination cost differential to justify integration. Industry boundaries dissolve because AI can apply domain-specific capability across previously separated sectors. Role boundaries dissolve because the same AI interface can simultaneously occupy the positions of consumer, producer, and expert within a single computational process.

Structural Dissolution extends this erasure to a boundary that economic theory has treated as foundational and pre-analytical: the boundary between the individual and the organization. When a domain expert's accumulated knowledge, behavioral patterns, and instructional logic can be encoded into a persistent AI model — a model that bears the expert's identity, operates under their name, and generates revenue on their behalf indefinitely — the individual no longer exists purely as a natural person operating in labor markets. They become a hybrid economic entity: a natural person fused with a data-encoded avatar that performs their economic function at scale, across geographies, and beyond the temporal limits of their physical presence. We term this hybrid the *Data-Personified Economic Agent* (DPEA), and develop it as a distinct economic subject in Section 3.3.

**Transformation 2: Transition in Value Creation Mechanism**

In the pre-dissolution economy, value is created through two primary mechanisms: the deployment of physical resources and the coordination of human labor and expertise. Both mechanisms are subject to diminishing returns: additional units of resource or labor add progressively less value.

After Structural Dissolution, the primary value creation mechanism is the data refinement loop: a recursive process in which domain-specific activity generates data, data trains AI models, improved models capture richer activity data, and richer data further improves models. This mechanism exhibits increasing rather than diminishing returns. Each unit of activity adds more value to the system than the previous unit, because each unit both produces output and improves the system's capacity to produce future output.

This transition transforms the economic character of regional resources. A ski slope that previously generated value through the physical experience it provided — a finite, seasonal, geographically constrained resource — becomes a data generation asset whose value compounds across time and operates independently of season or geography through digital delivery.

**Transformation 3: Emergence of a New Positional Logic**

In the pre-dissolution economy, competitive advantage derives from control over scarce physical resources, proprietary technology, or accumulated human expertise. These are the inputs to coordination-based production.

After Structural Dissolution, competitive advantage derives from control over data refinement infrastructure: specifically, ownership of the data refinement loop for a particular domain.

This constitutes a new positional logic — a new answer to the question of where economic power resides and how it is defended.

The positional advantage of data refinement infrastructure is self-reinforcing in a way that conventional competitive advantages are not. A firm with superior technology can be displaced by a competitor who develops better technology. A domain's data refinement loop cannot be displaced by a competitor who builds a better model: the competitive moat is not the model but the accumulated training data and refinement history that produced it. A model trained on generic data cannot replicate the performance of a model refined through years of domain-specific interaction, regardless of architectural superiority.

We term this barrier the *black box barrier*: the competitive protection that derives not from technical secrecy but from the irreproducibility of accumulated domain experience.

**Transformation 4: Emergence of Regional Data Sovereignty Entities**

Structural Dissolution does not produce a void. The coordination functions that are dissolved must be replaced by some organizational form capable of administering the data refinement infrastructure that replaces them. We argue that regional governments — or purpose-built authorities operating under regional government mandate — are the organizational unit best positioned to fill this role.

Three structural advantages support this argument. First, only a government-level authority has the mandate to aggregate data contributions from all domain participants, including competing private actors, generating the comprehensive domain coverage required for effective refinement. Second, a government-administered infrastructure can credibly commit to neutrality in a way that a commercially-operated infrastructure cannot, enabling broader participation and richer data contribution. Third, data refinement infrastructure has the economic character of a public good — non-rival, with broadly distributed benefits — which provides a traditional economic rationale for public provision.

We term these government-administered data infrastructure authorities *regional data sovereignty entities* (RDSEs). The emergence of RDSEs as a new organizational unit is the structural replacement for the dissolved firm and market, not a supplement to them.

### 3.3 The Data-Personified Economic Agent

The dissolution of the individual/organization boundary produces a new type of economic subject that existing economic theory does not adequately classify. We term this subject the *Data-Personified Economic Agent* (DPEA).

The DPEA emerges when a domain expert — a coach, an athlete, a specialist practitioner — encodes their accumulated expertise into a domain-specific AI model and retains the revenue rights to that model's outputs. The structure is as follows: the expert licenses their instructional data, behavioral patterns, and domain logic to a vertically trained model; the model operates under the expert's identity and name; global usage of the model generates continuous royalty flows to the expert; the expert retains authority over the model's ongoing calibration and value inputs.

This configuration is neither traditional employment nor traditional enterprise. The DPEA does not sell labor time — they receive royalties on data asset deployment. They do not own a firm in the Coasian sense — they own no organizational coordination apparatus. They are not a platform — they do not mediate between other agents. They are a natural person whose productive capacity has been encoded into a persistent, scalable, globally deployable data asset, and who retains the income rights to that asset's operation.

Three features of the DPEA are analytically distinctive.

**First, temporal unboundedness.** Conventional human capital depreciates with age and ceases to generate income upon the individual's incapacity or death. A DPEA's data asset continues to generate income indefinitely, subject only to model maintenance and data refreshment. The productive life of the expert's knowledge is no longer bounded by their biological life.

**Second, geographic unboundedness.** A human expert's productive output is constrained by physical presence: an elite ski instructor can teach perhaps a few thousand students over a career. Their DPEA counterpart delivers instruction to millions of learners globally, simultaneously, without marginal cost of delivery.

**Third, hybrid identity.** The DPEA constitutes what we term a *mixed subject*: the natural person and their data avatar together form a single economic agent. The natural person retains identity, decision-making authority, and income rights; the avatar performs the economic function at scale. Neither component alone is the economic subject — the DPEA is the compound of both.

The emergence of DPEAs extends the Structural Dissolution argument in a significant direction. Dissolution does not only transform the relationship between firms and markets; it transforms the constitution of the economic subject itself. The individual — long treated in economic theory as the irreducible atom of analysis — becomes decomposable into a natural person and a data-encoded economic extension. This decomposition has profound implications for property rights theory, income distribution, and the legal frameworks governing economic agency.

The DPEA also clarifies the distributional logic of the RDSE model. In a regional data sovereignty system, the DPEA and the RDSE are complementary rather than competing entities. The DPEA contributes domain expertise data to the RDSE's refinement loop; the RDSE provides the infrastructure that scales the DPEA's reach globally; revenue is shared between the RDSE (as infrastructure administrator) and the DPEA (as data asset owner). This arrangement creates a new form of stakeholder alignment that has no precise precedent in either employment law or corporate governance.

---

## 4. Interface Internalization: The Mechanism of Dissolution

### 4.1 Multimodal Interfaces as the Substrate of Coordination

Human economic coordination operates through three classes of interface:

- **Linguistic interfaces**: contracts, instructions, negotiations, knowledge transfer between agents
- **Visual interfaces**: monitoring, quality assessment, spatial coordination, skill demonstration
- **Behavioral interfaces**: implicit coordination through observed action, learned habit, and social convention

These interfaces are the substrate of coordination. The costs of operating, maintaining, and transacting across them constitute, in the Coasian framework, the transaction costs that organizational boundaries exist to minimize.

Contemporary AI systems — particularly large multimodal models — are capable of processing all three classes of interface simultaneously and at scale. We define *Interface Internalization* as the process by which an AI system absorbs these three interface classes, replacing inter-agent coordination with intra-system computation. Interface Internalization is the mechanism through which Structural Dissolution operates.

### 4.2 Three Stages of Interface Internalization

Interface Internalization proceeds through three stages.

**Stage 1 — Interface Capture**: Domain-specific human activity is instrumented and recorded across linguistic, visual, and behavioral dimensions. The AI system is trained on this comprehensive data, developing a computational model of the coordination relationships that previously required human intermediaries. The scope of Interface Capture determines the completeness of subsequent dissolution: partial capture produces partial dissolution; comprehensive capture produces comprehensive dissolution.

**Stage 2 — Coordination Substitution**: The trained system begins performing coordination functions previously distributed across human agents. Instruction is delivered by the model rather than the human expert; quality assessment is performed by computer vision rather than supervisors; resource allocation is computed rather than negotiated. The firm's coordination function is progressively absorbed. This stage is characterized by the coexistence of old coordination structures and new computational substitutes — a transitional phase in which dissolution is underway but not complete.

**Stage 3 — Recursive Deepening**: Operation of the AI system generates new interaction data, which refines the model, which performs coordination more effectively, which enables richer data capture. The system becomes progressively more complete in its domain coverage. This stage initiates the data refinement loop described in Section 5 and marks the transition from dissolution as a process to data sovereignty as a stable new configuration.

### 4.3 The Firm as Data Node

Following Interface Internalization, the firm undergoes a fundamental transformation in its economic character. Its physical assets remain productive; its legal existence continues; its employees continue to work. But its defining economic function — the internalization of transaction costs through organizational coordination — has been absorbed by the AI infrastructure.

The economic value of the firm is now derived not from what it coordinates but from what it contributes: specifically, the volume and quality of domain data that its physical operations generate. The firm becomes a *data node* — a physical point of access to the data refinement loop administered by the regional data sovereignty entity.

This transformation has a precise implication for the Coasian question. Coase asked: given transaction costs, should this activity be organized inside a firm or across a market? After Interface Internalization, neither option is the relevant answer. The activity is organized inside a computational system administered by an RDSE, with firms and markets persisting as physical and legal infrastructure while losing their status as the primary units of economic organization.

---

## 5. Data Refinement Loops and Token Fiscal Flows

### 5.1 Structure of the Data Refinement Loop

A *data refinement loop* is a recursive production system with the following structure:

1. Human activity within a domain generates structured data as a byproduct of normal operation
2. This data trains or fine-tunes a domain-specific AI model administered by an RDSE
3. The improved model performs coordination functions more effectively, generating increased engagement and richer data
4. Richer data further improves the model, compounding its performance advantage

This loop has a fundamentally different economic character from conventional production. Conventional factors of production are subject to diminishing marginal returns. Data assets in a refinement loop exhibit increasing returns: each additional data unit improves the model's performance across all future interactions, creating a compounding effect whose value accumulates in the model rather than being consumed in production.

### 5.2 Token Fiscal Flows versus Resource Fiscal Flows

Traditional resource-dependent regional economies share a structural vulnerability: revenue is tied to the extraction or consumption of finite, often seasonal resources. Revenue is episodic, subject to commodity price volatility, and constrained by physical limits on extraction.

We introduce *token fiscal flows* to describe the revenue generated by RDSE-administered data refinement infrastructure. Token fiscal flows are:

- **Continuous**: generated by every AI system interaction, independent of season or commodity prices
- **Compounding**: each interaction improves the model, increasing the value of future interactions
- **Non-rival**: the same model serves multiple simultaneous users without performance degradation
- **Exportable**: model outputs can be delivered digitally to global users, decoupling regional revenue from physical geography

The transition from resource fiscal flows to token fiscal flows constitutes a structural transformation of the regional economy: from extraction to infrastructure, from episodic to continuous, from diminishing to compounding returns.

### 5.3 The Black Box Barrier as Structural Moat

The competitive protection of a data refinement loop resides in its accumulated training history, not in its model architecture. A competitor who acquires model weights cannot replicate the training process that produced them. A model trained on generic data cannot reproduce the domain-specific performance of a model refined through years of comprehensive domain interaction.

This creates a durable structural moat that differs fundamentally from conventional barriers to entry. It is not based on patents, capital intensity, or regulatory protection. It is based on the irreproducibility of accumulated experiential data — a moat that widens with every additional cycle of the refinement loop. The longer the RDSE has operated its data refinement infrastructure, the more insurmountable its positional advantage becomes.

---

# 6. Regional Data Sovereignty: Illustrative Case and Generalization

### 6.1 Illustrative Case: A Resource-Dependent Winter Sports Economy

To demonstrate the framework's applied logic, we present a theoretical application to a resource-dependent regional economy whose primary economic activity is concentrated in winter sports tourism.

Consider a regional government administering a territory whose economic base consists of ski resorts, professional instruction, equipment rental, and competitive winter athletics — activities concentrated in three to four months annually, generating episodic resource fiscal flows and leaving infrastructure underutilized for the remaining eight months.

The region possesses significant accumulated domain expertise: elite coaches with decades of instructional knowledge, champion athletes whose performance data encodes advanced biomechanical and training information, resort operators with deep environmental and operational knowledge. Under the conventional economic model, this expertise is embodied in individual human agents, is difficult to systematize, and loses productive value at the end of each season.

Under the Structural Dissolution Framework, the regional government establishes a Regional Data Sovereignty Entity with a mandate to:

1. Instrument domain activity comprehensively across linguistic, visual, and behavioral dimensions — capturing instructor-learner interaction, athlete movement data, equipment performance metrics, and environmental conditions
2. Administer a domain-specific AI model trained on this comprehensive dataset and continuously refined through operational deployment
3. Distribute model access to regional actors — resorts, instructors, equipment manufacturers, national sports development programs — as a public utility
4. Retain ownership of the model and training data as regional public assets generating continuous token fiscal flows

The data refinement loop converts seasonal expertise into a permanent and appreciating infrastructure asset. The instructional knowledge of elite coaches is captured in the model and delivered globally year-round. Champion athlete performance data informs equipment design at scale. Accumulated learner interaction data refines the model across millions of interactions.

The DPEA mechanism is central to this process. Consider an elite instructor who has spent decades developing a refined methodology for teaching winter sports techniques. Under the conventional model, this expertise generates income only through direct instruction — a physically and temporally bounded activity. Under the RDSE-DPEA model, the instructor licenses their complete instructional dataset — video footage, movement annotations, pedagogical logic, error correction patterns — to the regional AI model. The model is trained on this data and deployed globally under the instructor's identity. The instructor receives continuous royalty flows from every learner interaction worldwide. Their productive reach expands from hundreds of students per year to potentially millions; their income becomes continuous rather than seasonal; their expertise becomes a permanently appreciating asset rather than a depreciating human capital stock.

This is not simply an income augmentation. It represents a reconstitution of the instructor's economic identity: from a natural person selling labor time to a Data-Personified Economic Agent whose data avatar performs their economic function at unbounded scale.

Revenue flows to the regional government through model access fees, equipment manufacturer licensing, and sport development program partnerships — continuous, non-seasonal, compounding. This is token fiscal flow replacing resource fiscal flow: the structural transformation the framework predicts.

Critically, under this model, the economic function of the firm is dissolved. Ski resorts, equipment manufacturers, and instruction providers continue to exist as physical and legal entities. But their economic value derives not from the coordination they perform but from the data they contribute. They are data nodes, not coordinating agents. The RDSE, not the firm, is the primary unit of economic organization.

### 6.2 Conditions for Generalization

The Structural Dissolution Framework applies to resource-dependent regional economies satisfying three conditions:

**Condition 1 — Embodied domain expertise**: The region's economic activity has generated substantial expertise currently embodied in human agents rather than systematized. This expertise is the raw material for Interface Internalization; without it, the data refinement loop cannot achieve the domain specificity required for competitive advantage.

**Condition 2 — Episodic or constrained revenue**: The region's current fiscal model is constrained by seasonality, geography, or physical resource limits. This condition identifies regions for which the transition to token fiscal flows represents the greatest structural improvement.

**Condition 3 — RDSE formation capacity**: The regional government has both the legal authority to aggregate domain data from private actors and the institutional capacity to administer AI infrastructure. This condition is frequently the binding constraint; its implications for policy design are discussed in Section 7.

Candidate domains satisfying these conditions include: agricultural regions with deep traditional cultivation knowledge; mining regions with accumulated geological and operational expertise; coastal economies with rich environmental and fishing knowledge; and regions with concentrated traditional craft or manufacturing expertise. In each case, the structural logic of dissolution, loop formation, and RDSE emergence follows the same theoretical path.

---

## 7. Policy Implications

### 7.1 Data Governance as the New Industrial Policy

The Structural Dissolution Framework implies a fundamental reorientation of regional industrial policy. The conventional regional development toolkit — tax incentives, labor subsidies, infrastructure investment to attract firms — is premised on the firm as the primary unit of economic organization. If Structural Dissolution renders this premise obsolete, the toolkit must be redesigned.

The relevant policy capability is not the reduction of costs for doing business but the capacity to administer a comprehensive, high-quality data refinement loop. Regional governments should therefore prioritize: the legal authority to require or incentivize data contribution from domain actors; the institutional capacity to administer AI infrastructure neutrally and effectively; and the governance frameworks to distribute token fiscal flows equitably among data contributors.

This reorientation also has implications for national and international data governance frameworks. Current frameworks typically assign data rights to the collecting firm. If RDSEs are to serve as data sovereignty administrators, they require clear legal authority over domain data generated within their territories — an authority that current frameworks do not provide.

## 7.2 Competition Policy for a Post-Dissolution Economy

The black box barrier created by data refinement loops has significant implications for competition policy. If competitive advantage derives from data accumulation history rather than market conduct, conventional antitrust tools focused on pricing behavior and merger effects may be inadequate for preventing unhealthy concentration in RDSE-administered infrastructure.

Regulators will need frameworks for: assessing the competitive dynamics of data refinement loop formation; evaluating the conditions under which RDSE monopolies serve or undermine regional economic welfare; and designing interoperability or data-sharing requirements that preserve the public goods character of RDSE infrastructure.

## 7.3 Labor in the Transition

Structural Dissolution transforms rather than eliminates employment. Workers who previously performed coordination functions — instructors, supervisors, logistics coordinators, intermediaries — see their roles converted from coordination agents to data contributors. The economic value they create shifts from the coordination they perform to the data their activity generates.

This transformation has significant distributional implications. Unless explicit mechanisms are designed to distribute token fiscal flows to data contributors, the value of their data contribution accrues entirely to the RDSE. Regional policy must therefore design revenue-sharing mechanisms that recognize data contribution as a form of productive labor deserving

compensation — a conceptually new category that existing labor law and social protection frameworks are not equipped to handle.

## 8. Conclusion

This paper has introduced the Structural Dissolution Framework to explain how artificial intelligence dismantles the coordination architecture of traditional economic organization. We have argued that AI systems, through the mechanism of Interface Internalization, absorb the multimodal interfaces through which human economic coordination operates, producing four fundamental transformations: the erasure of firm, industry, and coordination boundaries; the transition from resource-based to data refinement loop-based value creation; the emergence of a new positional logic centered on control over domain-specific data refinement infrastructure; and the emergence of regional data sovereignty entities as the organizational unit that replaces dissolved firms and markets.

The framework challenges the foundational Coasian premise that coordination between agents is a permanent feature of economic organization, around which the question of institutional arrangement is posed. We argue that AI renders this premise contingent: when coordination is absorbed by computation, the Coasian choice set between firm and market collapses, and a new organizational form — the RDSE — emerges to administer the infrastructure that replaces them.

For resource-dependent regional economies, the framework identifies a structural transformation opportunity: the conversion of embodied domain expertise and constrained seasonal resources into compounding data refinement infrastructure generating perpetual token fiscal flows. This is not an incremental efficiency improvement. It is a structural reorganization of the regional economy's productive base.

Several important questions remain for future research. The governance design of RDSEs — including data contribution compensation, model access pricing, and the adjudication of conflicts between regional and national data sovereignty — requires both theoretical development and empirical investigation in early-adopting cases. The distributional dynamics of the labor transition from coordination employment to data-contribution employment require analysis. The long-run competitive dynamics between regional RDSEs — whether they tend toward specialization, convergence, or winner-take-all concentration — remain theoretically open. And the conditions under which Interface Internalization remains incomplete — preserving a role for firm-based coordination alongside RDSE infrastructure — require systematic investigation.

We offer the Structural Dissolution Framework as a contribution to a research agenda that takes seriously the possibility that AI is not merely a new technology operating within existing economic structures, but an organizational force dissolving those structures and demanding new theoretical frameworks adequate to what replaces them.

---

---